\documentclass[twocolumn]{revtex4}
\usepackage[dvips]{graphicx}
\def\be{\begin{equation}}
\def\ee{\end{equation}}
\def\bea{\begin{eqnarray}}
\def\eea{\end{eqnarray}}

\begin{document}
\title{Study of COVID-19 epidemiological evolution in India with a multi-wave SIR model}
\author{Kalpita Ghosh}\email{kalpita_das@yahoo.co.in}
\affiliation {Charuchandra College, 22 Lake Place Road, Kolkata 700029, India}
\author{Asim Kumar Ghosh}
 \email{asimk.ghosh@jadavpuruniversity.in}
\affiliation {Department of Physics, Jadavpur University, 
188 Raja Subodh Chandra Mallik Road, Kolkata 700032, India}
\begin{abstract}
  The global pandemic due to the outbreak of COVID-19
  ravages the whole world  for more than two years
  in which all the countries are suffering a lot since December 2019.
  In order to control this ongoing  waves of epidemiological
  infections, attempts have been made to understand the
  dynamics of this pandemic in deterministic
  approach with the help of several mathematical models.
  In this article characteristics of a multi-wave SIR model
  have been studied which successfully explains the features of
  this pandemic waves in India.
  Stability of this model has been
  studied by identifying the equilibrium points as well as
  by finding the eigen values of the corresponding Jacobian matrices.
  Complex eigen values are found which ultimately
  give rise to the oscillatory solutions for the three
  categories of
  populations, say, 
  susceptible, infected and removed.
  In this model, a finite probability of the recovered people
  for becoming susceptible again is introduced which eventually lead to
  the oscillatory solution in other words. The set of differential
  equations has been solved numerically in order to obtain the
  variation for numbers of susceptible, infected and removed
  people with time. In this phenomenological study, finally
  an additional modification is made
  in order to explain the aperiodic oscillation which is found necessary to
  capture the feature of epidemiological waves particularly in India.
\end{abstract}
\maketitle
\section{INTRODUCTION}
The outbreak of COVID-19 in December 2019,
due to the spreading of infectious corona virus
named SARS-CoV-2 finally triggered
 to a global pandemic in which
all the countries are suffering a lot till to date.
This highly contagious disease has been transmitted to
millions of people globally where a fraction of
infected people is succumbed to it eventually.
Most of the countries witness multiple peaks of epidemiological
infections in its evolution which is
counted as number of waves. Among the most affected
countries, for example, countries in
European union (CEU), USA, Russia and Canada
experience five successive epidemiological waves while the third
wave is going on in India, Indonesia and Brazil. Which means that five peaks
of the epidemiological infections are
found in CEU, USA, Russia and Canada, while 
three distinct peaks
are noted in India, Indonesia and Brazil.
However, peaks of epidemiological
infections in India are sharper than those found in Brazil. 
So, it is obvious that characteristics of these
epidemiological waves found in different
countries are not similar. Widths and heights of the peaks are
different as well as the separation between them
are not equal.
It is observed that the same
individual gets infected multiple time during this pandemic
in all countries. 

In order to study the dynamics of the pandemic in a deterministic
approach, several models
have been proposed based on the classic SIR model introduced by
Kermack and Mckendrick in 1927. \cite{SIR} Most of the models
are mainly formulated to study the characteristics of the
first wave of the pandemic. Those are known as SEIR, SIQR and their
hybrids which are all derived from the single-wave SIR (SWSIR) model.
\cite{Hethcote,Dickmann-Hethcote,Keeling-Rohani,Martcheva,Daley-Gani}
It is termed as SWSIR model because of the fact
that this primitive version in general 
yields a solitary peak in its evolution. 
In those attempts, effects of quarantine,
isolation, latent time of infection
and other factors are taken into account where
several predictions and forecasts are available.
\cite{Hethcote-Zhien-Shengbing,Jumpen-Wiwatanapataphee-Wu-Tang}
Nonetheless,
periodic outbreak of disease in terms of SIR, SIQR, and SEIQR models are
studied analytically in great details. 
\cite{Hethcote-Levin,Feng-Thieme1,Feng-Thieme2,Feng-Thieme3}
Evolution of the COVID-19 pandemic for different countries have been studied
with the help of those models where origin of 
differences in its features are explained. Case studies on few
countries are available.
\cite{China,Italy,NewZealand,Brazil,France,UK-Israel,Canada}

\begin{figure}[h]
    \centering
  \includegraphics[width=\columnwidth]{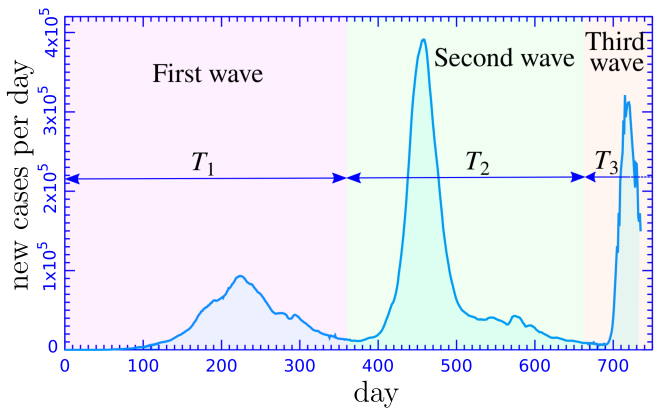}
  \caption{Daily new COVID-19 cases in terms of seven-day rolling average in India
    during January 30, 2020 to February 3, 2022 (736 days). The regions
    for first, second and ongoing third waves are
    demarcated by different colors.}
   \label{covid-india}
  \end{figure}

Hence, a multi-wave SIR
(MWSIR) model is formulated by modifying the
SWSIR model in order to explain the
dynamics of epidemiological infections including the
origin of multiple waves found in the infection pattern.
This is a fact that the same individual has been infected repeatedly
during this pandemic period. 
So in this model, probability of the same individual
for getting infected multiple times after curing
is taken into account.\cite{SODE1}
Time delay between successive infections of the same individual
is introduced in the MWSIR model. In this work,
periodic outbreak of COVID-19 in India will be studied, and so, the
daily new cases in terms of seven-day rolling average in India
during January 30, 2020 to February 3, 2022 (736 days)
are shown in Figure \ref{covid-india}.
  It reveals that the duration of those waves are different, so, it is not
  perfectly periodic. The approximate duration of first and second waves are
  $T_1=370$ and $T_2=300$ days, respectively, while the third wave crosses
  66 days ($T_3>66$) by this time.
  Hence, separation between adjacent peaks are different.
  In addition to that width and height of the peaks are unequal.
 
  In the section \ref{MWSIR} a MWSIR
  model is considered which gives rise to almost periodic
  epidemiological oscillation. 
  The stability of this model is studied
  in section \ref{stability} analytically while its periodic feature
  is studied numerically in section \ref{numerical}.
  In order to produce aperiodic oscillation, this model has
  been further modified. The epidemiological oscillation
  produced by this model is compared with the daily new COVID-19 cases
  in India. 
A discussion based on those results is available 
in section \ref{Discussion}.
\section{Multi-wave SIR model}
\label{MWSIR}
In order to explain the epidemiological waves of
infection due to the spreading of COVID-19
in India, a multi-wave SIR model is formulated
where the total population is divided into three dynamic sub-populations,
those are known as susceptible, infected and removed.
Susceptible population is constituted by the individuals
those are otherwise healthy but have a 
probability for having infected at any time in future.
The intermediate stage is comprised of infected population where
the individual is instantaneously
contagious. In the terminal stage individuals are either recovered or
succumbed to the disease, however, jointly referred to as
removed population. As the ongoing COVID-19 pandemic
exhibits periodic outbreak of the disease globally, a new
model is necessary to study the dynamics of this pandemic.

\begin{figure}[h]
    \centering
  \includegraphics[width=\columnwidth]{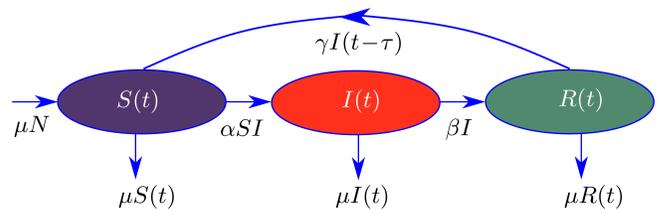}
  \caption{Flowchart for the MWSIR model.}
   \label{sir-flowchart}
  \end{figure}

Therefore, in order to understand the nature of periodic outbreak, 
a MWSIR model has been formulated by modifying the
primitive version of SIR model as described below. In this model the
recovered population has a nonzero probability to get infected again.
As a result, the susceptible population gets enriched with time
which is measured
in terms of infected population with a time delay, $\tau$. Thus,
the multi-wave SIR model
is defined by a set of coupled first-order nonlinear differential
equations (Eq \ref{sir}):
\begin{equation}
\left\{\begin{array}{l}\!\frac{d { S}}{dt}=\mu N-\alpha\, { S(t)}\,{ I(t)}
   +\gamma \,I(t\!-\!\tau)-\mu S(t), \\[0.5em]
  \! \frac{d {I}}{dt}=\alpha\, { S(t)}\,{ I(t)}-\beta\, { I(t)}-\mu I(t) ,
   \\[0.5em]
  \! \frac{d { R}}{dt}=\beta\,{ I(t)}- \gamma\, I(t\!-\!\tau)-\mu R(t),
 \end{array}\right. 
 \label{sir}
\end{equation}
where
$S(t)$, $I(t)$, and $R(t)$ are the number of
susceptible, infected and removed people at time $t$.
The positive constants $\mu, \alpha, \beta$ and $\gamma$
are the rates of birth per individual, infection,
removed and susceptibility,
respectively. The model incorporates the fact that 
the recovered people become susceptible again after the
mean period $\tau$.\cite{SODE1} However, $I(t\!-\!\tau)=0$, when
$t<\tau$. The SWSIR model will be restored when $\mu=0$, and $\gamma =0$.
The system in this formulation is closed in a sense that 
the total population, $N=S(t)+I(t)+R(t)$, 
does not change with time, which means the rate at which individual
suffers natural mortality is also given by the parameter
$\mu$. In other words, mean lifespan of individual turns out to be $1/\mu$.
In the same way, $1/\beta$ can be regarded as the mean recovery time for
the infected individual. 
The flowchart of this model is shown in Figure \ref{sir-flowchart}.

The basic reproduction number, 
$R_0$ plays an important role in epidemiology,
which is defined as average number of new infections per
infected individual. So, in this three-compartment
system, it is equal to the transmission rate
multiplied by the infectious period, $R_0=\alpha/(\beta+\mu)$.
\cite{Keeling-Rohani,Diekmann-Heesterbeek-Roberts}
The value of $R_0$ is crucial to have
an idea about how does the disease flow in the whole population and
 at the same time it provides clue to control its spreading.

  \section{Stability analysis of multi-wave SIR model}
  \label{stability}
  In order to determine the stable equilibrium points, the rate of change of
  $S(t)$, $I(t)$ and $R(t)$ with time are made zero. The trivial solution
  leads to $I(t)=0$ at any time, which corresponds to disease-free 
  equilibrium (DFE). Mathematically, this point is expressed as
  $(S^*,I^*,R^*)_{\rm tr}=(N,0,0)$, where the entire population
  becomes susceptible but no infected individual. This is true
  for both $t<\tau$ and $t\ge\tau$.  Anyway,
  this DFE is insignificant as it does not
  correspond to dynamics of epidemic by any means.

  The nontrivial solution leads to $I(t)\ne 0$, at finite time,
  but, $I(t\rightarrow \infty)\rightarrow  0$, which
  corresponds to endemic equilibrium and that
  is observed in SWSIR model. However, in this MWSIR model,
  nontrivial solutions mean $I(t)> 0$, at any time. So these 
  dynamic equilibrium points appear periodically with time
  without showing any signature of ending if there is no damping.
  These equilibrium points
  are marked by $(S^*,I^*,R^*)_{\rm ntr}=\left(\frac{\beta+\mu}{\alpha},
  \frac{\mu N}{\beta+\mu}-\frac{\mu}{\alpha},
  \frac{\beta N}{\beta+\mu}-\frac{\beta+\mu}{\alpha}
  +\frac{\mu}{\alpha}\right)$, when $t<\tau$,
  however, for  $t\ge \tau$, 
   $(S^*,I^*(t),R^*(t))_{\rm ntr}=\left(\frac{\beta+\mu}{\alpha}, \frac{\mu N}{\beta+\mu}+\frac{\gamma I(t\!-\!\tau)}{\beta+\mu}-\frac{\mu}{\alpha},
  \frac{\beta N}{\beta+\mu}-\frac{\beta+\mu}{\alpha}
  -\frac{\gamma I(t\!-\!\tau)}{\beta+\mu}+\frac{\mu}{\alpha}\right)$.
  Hence, equilibrium points are static when $t<\tau$, but dynamic when 
$t\ge \tau$.

  In order to understand the dynamics close to the equilibria,
  Jacobian matrix ($J$) is constructed which can be expressed at
  the respective 
  equilibrium points. Jacobian matrix has the form
  \begin{equation}
    \left[\begin{array}{ccc}
        -(\alpha I^*+\mu) & -\alpha S^* &0
        \\[0.5em]
  \alpha I^* &\alpha S^* -\beta -\mu&0
   \\[0.5em]
 0&\beta & -\mu
 \end{array}\right]. 
\end{equation}
  The eigenvalues of Jacobian matrix ($\lambda_0,\lambda_\pm$) can be given as
\begin{equation}
    \left\{\begin{array}{l}\!\lambda_0=-\mu, \\[0.5em]
   \! \lambda_\pm=-\frac{\alpha (I^*-S^*)+\beta+2\mu}{2}\pm
    \frac{\sqrt{\alpha^2 (I^*-S^*)^2+\beta^2-2\alpha \beta (S^*+I^*)}}{2}.
 \end{array}\right. 
\end{equation}
For the DFE, $(S^*,I^*,R^*)_{\rm tr}=(N,0,0)$, the
eigenvalues are 
\begin{equation}
    \left\{\begin{array}{l}\lambda_0=-\mu, \\[0.5em]
     \lambda_+=\alpha N -\beta -\mu, \\[0.5em]
    \lambda_-=-\mu.
 \end{array}\right. 
\end{equation}
As a result, it will behave like a stable point
when $\alpha<(\beta+\mu)/N$, which is similar to
the SWSIR model. It is obvious that the eigenvalues
will be complex for the general case when 
\[\alpha \beta>\frac{\alpha^2 (I^*-S^*)^2+\beta^2}{2(S^*+I^*)}.\]
It is expected that variations of $S(t)$, $I(t)$, and $R(t)$
will be oscillatory around these points.
\section{Numerical results}
\label{numerical}
\begin{figure}[h]
    \centering
  \includegraphics[width=\columnwidth]{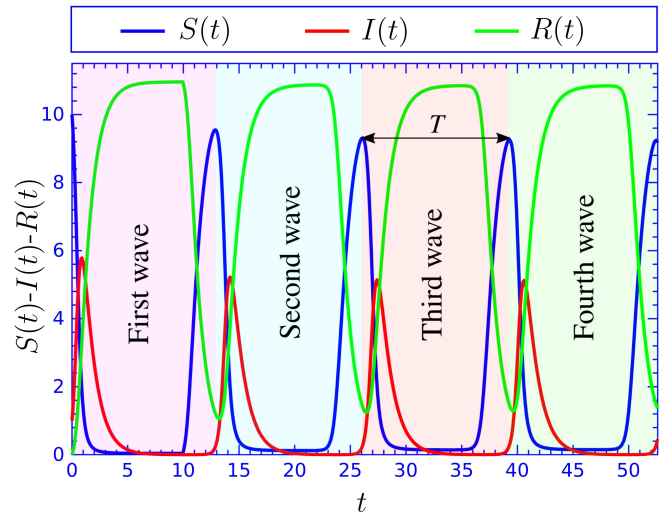}
  \caption{Four successive waves are shown here those are
    generated numerically by solving MWSIR model (Eq \ref{sir}).
    Values of the parameters: $S(0)=10,\,I(0)=1,\,R(0)=0,\,N=11,\,\alpha=0.5,\, \beta= 1,\, \gamma=1, \,
    \tau = 10,\,\mu=0.$
  Time period of oscillation is marked by $T$.}
   \label{sir-multiwave}
  \end{figure}

As the values of $S(t)$, $I(t)$, and $R(t)$ 
cannot be determined analytically by solving the set of equations,
Eq \ref{sir}, they have been solved numerically.
The oscillations of $S(t)$, $I(t)$, and $R(t)$ are
noted for long time and it is observed that
all are oscillating with almost constant amplitude. 
Four successive waves are shown in Fig \ref{sir-multiwave}, where it
is observed that the mean time period of oscillations, $T$,
is always greater than $\tau$.
As no damping factor is introduced here, amplitudes of the
respective waves for  $S(t)$, $I(t)$, and $R(t)$
do not change with time. 

\begin{figure}[h]
    \centering
  \includegraphics[width=\columnwidth]{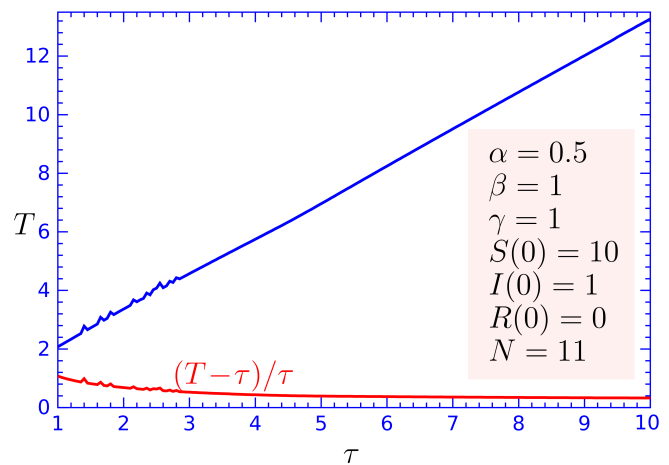}
  \caption{Variation of $T$ and $(T\!-\!\tau)/\tau$ with $\tau$, 
   in the MWSIR model for $S(0)=10,\,I(0)=1,\,R(0)=0,\,N=11,\,\alpha=0.5,\, \beta= 1,\, \gamma=1, \,
    \mu=0.$}
   \label{T-vs-tau}
 \end{figure}

The time period of oscillations for $S(t)$, $I(t)$, and $R(t)$
are obviously the same, although it 
decreases with the number of cycles. 
However the difference, $T\!-\!\tau$
vanishes with the increase of $\tau$.
This feature is shown in Fig \ref{T-vs-tau} where
variations of time period $T$ along with relative difference
between $T$ and $\tau$ or $(T\!-\!\tau)/\tau$ with $\tau$ are
plotted. 

 \begin{figure}[h]
    \centering
  \includegraphics[width=\columnwidth]{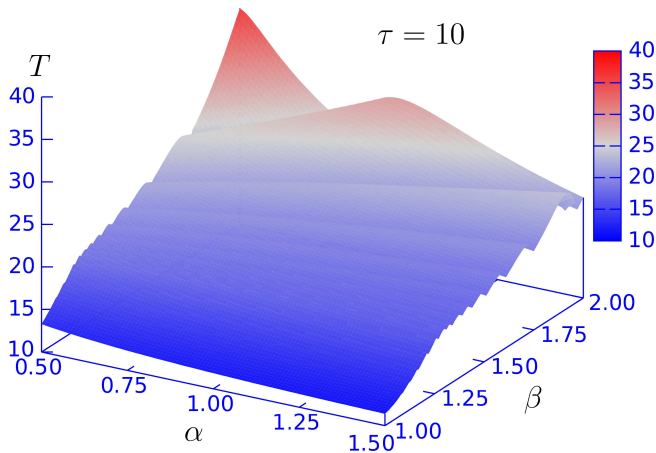}
  \caption{Variation of $T$ with $\alpha$ and $\beta$, when $\tau=10$, 
   in the MWSIR model for $S(0)=10,\,I(0)=1,\,R(0)=0,\,N=11,\, \gamma=1, \,
    \mu=0.$}
   \label{sir-variation-mena-time-with-rates}
 \end{figure}
 The variation of $T$ with $\alpha$ and $\beta$ for fixed value of
 $\tau$ is shown in Figure \ref{sir-variation-mena-time-with-rates}.
 It clearly indicates that $T>\tau$, for any values of $\alpha$ and $\beta$.
 Moreover, the difference, $T\!-\!\tau$ steadily decreases with
 $\alpha$ but increases gradually with $\beta$. 
 \begin{figure}[h]
    \centering
  \includegraphics[width=\columnwidth]{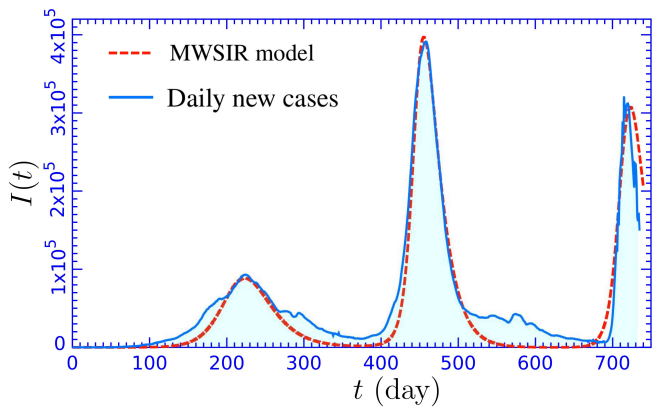}
  \caption{Daily new COVID-19 cases in terms of seven-day rolling average in India
    during January 30, 2020 to February 3, 2022 (736 days)
    is compared with the numerical results
    obtained by the MWSIR model. Values of the parameters are
    $\alpha=1.78\times 10^{-7},\,\beta=1/14,\,
    S(0)=7.3\times10^{5},\,\,I(0)=1,\,R(0)=0.$}
   \label{multiwave-covid-india}
 \end{figure}
 
Features of the ongoing COVID-19 epidemiological waves in India 
are shown in Figure \ref{covid-india}.
It reveals that
the duration of successive waves are decreasing as $T_1>T_2>T_3$.
On the other hand, 
amplitude of the second wave is the maximum while the third wave
is expected to cross its peak by this time. These features suggest that
the prevailing waves are not periodic by any means. Hence
the MWSIR model is further modified in order to 
explain this aperiodic behavior.
One can remember that transformation of SWSIR to MWSIR
is accomplished by introducing two parameters
$\gamma$ and $\tau$, where $\gamma$ is the rate at which
the recovered people become susceptible again, while $\tau$
is the time delay to estimate the number of susceptible people
in terms of infected people.
 
In order to capture this specific aperiodic features
in the most simple way, the value of $\gamma$ has been
split up for three different waves in the MWSIR model as
\[\gamma=\left\{\begin{array}{lc}0,& t<\tau, \\[0.3em]
 \gamma_1,& \tau \le t<\nu\,  \tau, \\[0.3em]
 \gamma_2,& t\ge \nu\,  \tau, 
\end{array}\right. \]
where $\gamma_1=0.166$, $\gamma_2=0.056$, $\nu=2.78$, and $\tau=175$ days.
The relation, $\gamma_2<\gamma_1$, corresponds to the fact that
height of third peak is lower than the second one.
The value, $\beta=1/14$, is kept fixed as it 
corresponds to the mean recovery period of COVID-19, $1/\beta=14$ days. 
The numerical data obtained by solving the MWSIR model
is plotted in red dashed line and compared with the
daily new COVID-19 cases (seven-day rolling average) in India
shown in blue sold line. A very good agreement is found when the
rate of infection is assumed very small ($\alpha=1.78\times 10^{-7}$).

This phenomenological study predicts that
16.6\% of individuals infected during January 30, 2020 to
December 5, 2020 (first wave) become susceptible again for
further infection. Similarly, 5.6\% of individuals infected during 
December 5, 2020 to May, 27, 2021 (second wave) become
susceptible again for the next infection. This reduction may 
account the effects of quarantine, isolation, vaccination
and other preventive measures. However, as the solutions of the
non-linear equations are highly sensitive to the initial
conditions, like $S(0),\,I(0)$, and $R(0)$, estimated values
of the parameters, $\alpha,\,\gamma,\,\nu$ and $\tau$ are likely
to change dramatically with the change of initial values.
Which on the other hand will change the
predictions for obvious reason.

\section{Discussion}
\label{Discussion}
In order to study the behaviour of ongoing pandemic of COVID-19 in India
mathematically, a modified version of
SWSIR model termed as MWSIR model is proposed here
which is found successful to
reproduce the features of the multiple epidemiological waves
observed in this country. The model is deigned in such a fashion that
it is able to yield periodic as well as aperiodic epidemiological waves
with varying widths and heights
by tuning its parameters. As the SARS-CoV-2 virus
responsible for the COVID-19 is highly infectious, the spreading of this
disease cannot be controlled easily by means of 
simple preventive measures like quarantine, isolation, lock down
and even vaccination. As a result, in due course,
a sizable fraction of recovered people
gets infected multiple times leading to multiple waves of the epidemiological
infection. In order to explain this particular feature, the SIR model has been
modified in such a manner that the recovered people has a finite
chance for becoming susceptible
again after their recovery in this modified version.
Estimated values of the parameters obtained in this study will
be useful for imposing restrictions to control the pandemic. 
It will become helpful even for making accurate forecast and prediction. 

However, in this simple model no time delay for the incubation period
of the pathogens within the human body 
is considered, and the effect of quarantine, isolation, vaccination  and other
macroscopic measures are not take into account. So, it is expected that 
more accurate prediction can be made with the help
of this model by accommodating
the effect of quarantine, isolation, vaccination,
lock-down and other factors in its modified
version.
Which means that the discrepancy between theoretical
estimation and daily new cases can be removed further
by formulating multi-wave SEIR, SIQR, SEIQR and other hybrid models. 
Obviously, this model can be employed
for examining the characteristics of epidemiological waves
found in other countries as well.

\end{document}